\documentclass[pra,preprint,tightenlines,showpacs,12pt]{revtex4} 
\usepackage{epsfig} 
\usepackage{color}

\begin{document} 

\title{ Symmetric eikonal model for projectile-electron excitation 
and loss in relativistic ion-atom collisions } 

\author{ A.B.Voitkiv and B.Najjari }   
\affiliation{ Max-Planck-Institut f\"ur Kernphysik, 
Saupfercheckweg 1, D-69117 Heidelberg, Germany } 
\author{ V.P.Shevelko } 
\affiliation{ P.N.Lebedev Physical Institute of the Russian Academy of Sciences, 
Leninsky Avenue 53, 117924 Moscow, Russia }


\begin{abstract} 

At impact energies $ \stackrel{>}{\sim}1$ GeV/u 
the projectile-electron excitation and loss occurring in 
collisions between highly charged ions 
and neutral atoms is already strongly influenced  
by the presence of atomic electrons. 
In order to treat these processes in collisions with heavy atoms 
we generalize the symmetric eikonal model, used earlier 
for considerations of electron transitions in ion-atom collisions 
within the scope of a three-body Coulomb problem. 
We show that at asymptotically 
high collision energies this model leads to an 
exact transition amplitude and is very well suited to 
describe the projectile-electron excitation and loss 
at energies above a few GeV/u. In particular, 
by considering a number 
of examples we demonstrate advantages of this model over 
the first Born approximation at impact energies  
$\sim 1$--$30$ GeV/u, which are of special interest 
for atomic physics experiments at the future GSI facilities.   

\end{abstract}  


\maketitle 


\section{Introduction} 

During the last two decades there has been 
performed a large number of experimental investigations 
of projectile-electron excitation and loss 
in collisions between relativistic 
highly charged ions and solid and gaseous targets.  
In particular, a variety of very heavy projectiles 
with the net charge $52$ -- $91$ a.u.  
were used in the experiments.  
The experiments have also covered 
a very large interval of impact energies ranging from 
comparatively low relativistic energies 
of $\sim 100$-$200$ MeV/u \cite{standf}-\cite{low-middle-3}  
to extreme relativistic energy of $160$ GeV/u 
\cite{high-2}-\cite{high-4} where the projectile velocity 
already only fractionally differs from the speed of light $c=137$ a.u..   

Most of the data, however, have been collected 
for impact energies not exceeding a few hundreds of MeV/u.   
For impact energies above $1$ GeV/u  
there exists just a few experimental results. 
They include the data on the electron loss from  
$10.8$ GeV/u Au$^{78+}$(1s) ions penetrating 
solid targets \cite{high-0}-\cite{high-1} and 
on the electron loss from $160$ GeV/u Pb$^{81+}$(1s) ions 
colliding with solid \cite{high-2}
and gaseous \cite{high-3} targets. 

Besides, one should also note that 
the experimental data on the elementary 
cross sections in collisions at 
very high energies collected using solid targets 
are not very accurate. 
The loss cross sections, reported for $160$ GeV/u Pb$^{81+}$(1s) 
projectiles penetrating solid and gas targets 
differ between themselves roughly by a factor of $2$ 
(see \cite{high-2}-\cite{high-3}). 
The reason for this difference, 
as was recently explained in \cite{we}, 
lies in multiple collisions suffered  
by the projectiles when they move 
in solids which does not allow one 
an accurate experimental determination of the 
values for elementary ion-atom 
cross sections. In the case of $10.8$ GeV/u Au$^{78+}$(1s) ions 
the data of \cite{high-0} and \cite{high-1}, 
which were both collected for collisions with 
solid state targets, also differ by about of a factor of $2$ 
(but the reason for this is not completely clear). 

Concerning difficulties in the theoretical description 
of the ion-atom collisions, there should be mentioned 
two main points, which complicate the treatment 
of the projectile-electron excitation and loss:  
The presence of atomic electrons and the fact that 
in case of collisions with a heavy atom 
the atomic field may be too strong 
making it necessary to go beyond 
the first Born approximation.  

The latter point is known to be of extreme importance when  
considering the projectile-electron excitation and loss 
processes occurring in collisions of very highly 
charged ions with heavy atoms at relatively low impact energies 
$0.1$ -- $0.2$ GeV/u, where the difference between 
experimental data and first Born calculations reaches an order of magnitude. 
At such energies, however, the atomic electrons play only a minor role 
in the projectile-electron transitions and their presence can simply 
be neglected \cite{abv-book}.  

With increase in the impact energy 
the role of the higher order effects in the projectile-target interaction 
diminishes. Nevertheless, even at energies  
well above $1$ GeV/u the first Born approximation tends 
to substantially overestimate transition probabilities and 
the accuracy of first Born results for cross sections 
remains unclear, especially in the case when 
the collision causes more than one electron 
of the projectile to undergo transitions. 
Besides, at such impact energies 
even for the most highly charged projectiles 
(like e.g. hydrogen-like uranium ions) 
the influence of the electrons of the atom on 
the projectile-electron transitions can 
no longer be ignored \cite{abv-book}.   
  
In the present paper we shall demonstrate 
that the so called symmetric eikonal (SE) model, 
extended to account for the presence of atomic electrons,   
can be used for treating the projectile-electron excitation 
and loss in collisions with heavy atoms at impact 
energies $\stackrel{>}{\sim} 1$ GeV/u. 
In particular, we will show that in the limit 
of asymptotically high impact energies 
the transition amplitude, derived in the SE model,  
coincides with the amplitude obtained in the light cone approximation 
which means that the SE model provides essentially an exact solution 
for the problem of projectile-electron 
excitation and loss in extreme relativistic ion-atom collisions.  

The forthcoming upgrade of 
the heavy-ion facilities  
at GSI (Darmstadt, Germany) 
will allow one to perform extensive 
experimental explorations of 
the different aspects of 
heavy ion -- atom collisions 
at impact energies in the range $ 1 $ -- $ 30 $ GeV/u. 
It will be seen below that in this range 
of impact energies the SE model represents    
a very valuable tool for describing 
the projectile-electron excitation and  
loss in collisions with very heavy atoms.   

Atomic units are used throughout unless otherwise stated. 

\section{ Theory }   

Depending on whether or not the electrons of the atom are 'active' 
in the collision, one normally distinguishes two atomic modes 
(see e.g. \cite{mont-review}, \cite{abv-book}) 
which can contribute to the projectile-electron transitions.  
In one of them, which is often called {\it screening},  
the electrons of the atom act coherently 
with the atomic nucleus screening (partially or fully) 
the field of the latter. The other mode, in which   
the atomic electrons actively participate undergoing 
transitions, is termed {\it antiscreening}. In this mode 
the behavior of the projectile-electron is influenced  
mainly by the electrons of the atom while 
the nucleus of the atom plays only a minor role \cite{ascr}.  

In collisions with heavy atoms, which are of special interest for 
the present article, the antiscreening mode is much weaker 
than the screening one and below will not be considered.  
In the screening mode the field of the atom can be regarded  
as external that enables one to reduce a many-electron problem 
of the ion-atom collision to a problem of the motion of the electron 
of the projectile in two external fields: the field of the nucleus 
of the ion and the field of the atom. The latter is taken as 
a superposition of the field of the atomic nucleus 
and the field of the atomic electrons whose space distribution 
is assumed to be 'frozen' during the very short collision 
time \cite{abv-book}.  

In the rest frame of the ion the electron is described by 
the Dirac equation  
\begin{eqnarray} 
i \frac{\partial \Psi({\bf r},t) }{ \partial t } = %
\hat{H} \Psi({\bf r},t).  
\label{e1}
\end{eqnarray}
The Hamiltonian $\hat{H}$ reads  
\begin{eqnarray}
\hat{H} = \hat{H}_0 +  \hat{W}, 
\label{e2}
\end{eqnarray}
where 
\begin{eqnarray}
\hat{H}_0 = c \mbox{\boldmath$\alpha$} \cdot {\bf p} %
+ \beta c^2 - \frac{Z_I}{r} 
\label{e3}
\end{eqnarray}
is the Hamiltonian for the electron motion 
in the field of the ionic nucleus and 
\begin{eqnarray}
\hat{W} = \mbox{\boldmath$\alpha$} \cdot {\bf A}({\bf r},t) %
 - \Phi({\bf r},t) 
\label{e4}
\end{eqnarray}
is the interaction between the electron 
and the field of the atom. 
In the above expressions 
$\mbox{\boldmath$\alpha$}$ and $\beta$ are the 
Dirac's matrices, ${\bf r}$ coordinates of the electron 
with respect to the ionic nucleus, $\Phi$ and 
$\bf{A}$ are, respectively, the scalar and vector 
potentials of the electromagnetic field 
of the incident atom.

The field of the atom in its rest frame 
is described by the scalar potential which,  
using results of \cite{Moliere} and \cite{Salvat}, 
can be taken as  
\begin{eqnarray} 
\Phi' = \frac{Z_A \phi(r')}{r'}, 
\label{e5}
\end{eqnarray}
where  
\begin{eqnarray} 
\phi(r') = \sum_{j} A_j \exp(-\kappa_j r') 
\label{e6}
\end{eqnarray} 
with the screening parameters  
$A_j$ ($\sum_j A_j= 1$) and $\kappa_j$ 
given in \cite{Moliere} and \cite{Salvat}. 
We assume that in the rest frame of the ion 
the atom moves along a classical 
straight-line trajectory  
${\bf R} = {\bf b} + {\bf v} t$,
where ${\bf b} =(b_x, b_y) $ 
is the impact parameter and  
${\bf v} =(0,0,v)$ the atomic velocity.   
Using Eqs.(\ref{e5})-(\ref{e6}) and 
the Lorentz transformation for the potentials 
we obtain that in the rest frame of the ion 
the potentials of the atomic field 
are given by   
\begin{eqnarray}
\Phi({\bf r},t) &=& %
\frac{\gamma Z_A}  
{\sqrt{\gamma^2(z-vt)^2+({\bf r}_{\perp} - {\bf b})^2}} %
\sum_{j} A_j \exp\left(-\kappa_j 
\sqrt{\gamma^2(z-vt)^2+({\bf r}_{\perp} - {\bf b})^2} \right).  
\nonumber \\ 
{\bf A}({\bf r},t) &=& \left(0,0, \frac{v}{c} \Phi \right) 
\label{e7} 
\end{eqnarray}
where $\gamma$ is the collisional Lorentz factor 
and ${\bf r} = ({\bf r}_{\perp}, z)$ with  
$ {\bf r}_{\perp} \cdot {\bf v} =0 $.  

Within the SE model  
the transition amplitude is approximated by  
\begin{eqnarray}
a_{fi}({\bf b}) = - i \int_{- \infty}^{+ \infty} dt %
\langle \chi_f(t) \mid %
\left( \hat{H}  - %
i \partial/\partial t \right) \chi_i(t) \rangle,   
\label{eic-new-1}  
\end{eqnarray}
where the initial and final states of the electron, 
whose motion in the field of the ionic nucleus 
is affected by the field of the atom,  
are chosen according to  
\begin{eqnarray}
\chi_i(t) &=& \psi_0 \exp(-i\varepsilon_0 t ) %
\exp\left( i \int_{-\infty}^{t} dt' \Phi(t') \right) 
\nonumber \\ 
\chi_f(t) &=& \psi_n \exp(-i\varepsilon_n t) %
\exp\left( i \int_{+\infty}^{t} dt' \Phi(t') \right).    
\label{eic-new-2} 
\end{eqnarray}
Here, $\psi_0$ and $\psi_n$ are the initial and final 
undistorted states of the electron in the ion. 

Making use of the fact that the dependence 
of the scalar potential $\Phi$ on 
the electron coordinates and time 
is of the form $ \Phi= \gamma Z_A f(s_{\perp}, \gamma |z - vt|) $, 
where ${\bf s}_{ \perp } = {\bf r}_{\perp} - {\bf b} $   
the transition amplitude (\ref{eic-new-1}) 
can be transformed into     
\begin{eqnarray}
a_{fi}({\bf b}) &=& i \frac{c}{v}\int_{- \infty}^{+ \infty} dt %
\exp(i \omega_{n0} t) %
\left\langle \psi_n \left|  %
\exp\left( i \int_{-\infty}^{+\infty} dt' \Phi(t') \right)
\times \right. \right.  
\nonumber \\ 
&& \left. \left.  
\left( \frac{\Phi(t)}{\gamma^2} \alpha_z - %
v \left( {\bf \nabla}_{\perp} \int_{-\infty}^{t} dt' \Phi(t') \right) \cdot %
\mbox{\boldmath$\alpha$}_{\perp} \right) \right| \psi_0 \right\rangle, 
\label{eic-new-3} 
\end{eqnarray}
where $\omega_{n0} = \varepsilon_n - \varepsilon_0$ 
is the electron transition frequency   
and ${\bf \nabla}_{\perp}$ denotes 
the two-dimensional (in the $(x,y)$-plane) 
gradient operator. 

The amplitude (\ref{eic-new-3}) 
can be simplified by employing the relation  
\begin{eqnarray}
&& \lim_{\lambda \to +0} \int_{-\infty}^{+\infty} dt  
\exp(i \omega_{n0} t) \exp( - \lambda |t|)  
\int_{-\infty}^{t} dt' \Phi(t')     
\nonumber \\ 
&& = \frac{ i }{ \omega_{n0} }
\int_{-\infty}^{+\infty} dt  
\exp(i \omega_{n0} t) \Phi(t),  \, \, (\omega_{n0} \neq 0),    
\label{eic-new-4}  
\end{eqnarray} 
that yields 
\begin{eqnarray} 
a_{fi}({\bf b}) &=& i \frac{c}{v}\int_{- \infty}^{+ \infty} dt %
\exp(i \omega_{n0} t) %
\left\langle \psi_n \left|  %
\exp\left( i \int_{-\infty}^{+\infty} dt' \Phi(t') \right)
\times \right. \right.  
\nonumber \\ 
&& \left. \left.  
\left( \frac{\Phi(t)}{\gamma^2} \alpha_z - %
i \frac{ v }{ \omega_{n0} } \left( {\bf \nabla}_{\perp} \Phi(t) \right) \cdot %
\mbox{\boldmath$\alpha$}_{\perp} \right) \right| \psi_0 \right\rangle.  
\label{eic-new-5} 
\end{eqnarray}

\subsection{ The Limit of weak interaction }

If the interaction between the electron of the ion and 
the atom is sufficiently weak one can replace the exponent 
$ \exp\left( i \int_{-\infty}^{+\infty} dt' \Phi(t') \right) $ 
in (\ref{eic-new-5}) by $1$. Then integrating in (\ref{eic-new-5})  
by parts over the space and using the continuity equation  
\begin{eqnarray}
\frac{\partial \rho_{n0}}{ \partial t} + 
{\bf \nabla}_{\perp} \cdot {\bf j}_{n0} =0 
\label{eik-1order-1}  
\end{eqnarray}
for the transition charge and current densities, 
\begin{eqnarray}
\rho_{n0} &=& \psi_n^{\dagger} \, \, \psi_0 \, \,  
\exp\left( i \omega_{n0} t \right) 
\nonumber \\ 
{\bf j}_{n0} &=& \psi_n^{\dagger}
\, \, c \mbox{\boldmath$\alpha$} \, \,  \psi_0 \, \,  
\exp\left( i \omega_{n0} t \right),
\label{eik-1order-2}   
\end{eqnarray} 
we obtain   
\begin{eqnarray}
a_{fi}({\bf b}) &=& i \frac{c}{v}\int_{- \infty}^{+ \infty} dt %
\exp(i \omega_{n0} t) %
\left\langle \psi_n \left| \frac{v}{c} \Phi  %
+ \alpha_z \left(\frac{\Phi}{\gamma^2} - 
i \frac{ v }{ \omega_{n0} } \frac{\partial \Phi}{ \partial z} \right) %
\right| \psi_0 \right\rangle.  
\label{eic-1order-3} 
\end{eqnarray}
Taking into account that 
$\frac{ \partial \Phi}{ \partial z } = - \frac{ 1 }{ v } \frac{ \partial \Phi}{ \partial t } $ 
and 
\begin{eqnarray}
\int_{-\infty}^{+\infty} dt \exp\left(i \omega_{n0} t \right) %
\frac{ \partial \Phi}{\partial t } = 
- i \omega_{n0} \int_{-\infty}^{+\infty} dt \exp\left(i \omega_{n0} t \right) \Phi  
\label{eic-1order-4} 
\end{eqnarray}
we arrive at the transition amplitude  
\begin{eqnarray}
a_{fi}({\bf b}) &=& i \int_{- \infty}^{+ \infty} dt %
\exp(i \omega_{n0} t) %
\left\langle \psi_n \left| \Phi \left( 1 - \frac{ v }{ c } \alpha_z \right)  %
\right| \psi_0 \right\rangle   
\label{eic-1order-5} 
\end{eqnarray}
which coincides with expression for the amplitude 
obtained in the first Born approximation.  

\subsection{ The High-Energy Limit }  

An important question, which will be 
addressed in this subsection, concerns 
the high-energy limit ($\gamma \to \infty$) 
of the SE model. Keeping in mind that 
$\Phi= \gamma Z_A f(s_{\perp}, \gamma |z - vt|)$ 
one can show that  
\begin{eqnarray}
\int_{-\infty}^{+\infty} dt  
\exp(i \omega_{n0} t) \Phi(t) &=&  
\frac{ 2Z_A }{ v } \exp\left(i \omega_{n0} z/v \right) 
\int_{-\infty}^{+\infty} d \xi f(s_{\perp}, \xi) 
\nonumber \\ 
&=& \exp\left(i \omega_{n0} z/v \right) \, 
G\left( s_{\perp}, \frac{\omega_{n0}}{\gamma v} \right), 
\label{h-e-1} 
\end{eqnarray}
where at the moment the explicit form of 
the function $G$ is not important. 
Correspondingly,  
\begin{eqnarray} 
\int_{-\infty}^{+\infty} dt  \, \, \Phi(t) = 
G\left( s_{\perp}, 0 \right) \equiv G_0\left( s_{\perp} \right).  
\label{h-e-2}  
\end{eqnarray} 
At sufficiently large impact energies, 
where the difference between 
$ G=G\left( s_{\perp}, \frac{\omega_{n0}}{\gamma v} \right) $ 
and $G_0$ essentially vanishes,  
we can replace the amplitude (\ref{eic-new-5}) 
by the following expression 
\begin{eqnarray} 
a_{fi}({\bf b}) &=& i \frac{c}{v \gamma^2}  
\left\langle \psi_n \left|  %
\exp\left( i G \right) %
\exp( i \omega_{n0} z/v ) \, %
G \, \alpha_z \right| \psi_0 \right\rangle %
\nonumber \\ 
&-& i \frac{c}{\omega_{n0}} 
\left\langle \psi_n \left|  %
\exp\left( i G \right) %
\exp( i \omega_{n0} z/v )  
\left( {\bf \nabla}_{\perp} G \right) \cdot %
\mbox{\boldmath$\alpha$}_{\perp}  
\right| \psi_0 \right\rangle.  
\label{h-e-3}  
\end{eqnarray}
We now take the second line 
of (\ref{h-e-3}), integrate there by parts, 
use the continuity equation (\ref{eik-1order-1}) and 
then again integrate by parts.  As a result 
of these manipulations, expression (\ref{h-e-3}) 
transforms into   
\begin{eqnarray} 
a_{fi}({\bf b}) &=& 
\left\langle \psi_n \left|  %
\exp( i \omega_{n0} z/v ) %
\exp\left( i G \right) %
\left( 1 - \frac{v}{c} \alpha_z \right) %
\right| \psi_0 \right\rangle %
\nonumber \\ 
&-& \frac{c}{v \gamma^2}  
\left\langle \psi_n \left|  %
\exp( i \omega_{n0} z/v ) 
\exp\left( i G \right) %
\left( 1 - i G \right) \alpha_z 
\right| \psi_0 \right\rangle.  
\label{h-e-4}  
\end{eqnarray}
The limit $\gamma \to \infty$ 
of the amplitude (\ref{h-e-4}) is given by 
\begin{eqnarray} 
a_{fi}({\bf b}) &=& 
\left\langle \psi_n \left|  %
\exp( i \omega_{n0} z/v ) %
\exp\left( i G_0 \right) %
\left( 1 - \frac{v}{c} \alpha_z \right) %
\right| \psi_0 \right\rangle   %
\label{h-e-5}   
\end{eqnarray}
and it coincides with 
the transition amplitude derived 
in the so called light-cone approach 
(see \cite{baltz}, \cite{abv-book}). 

The light-cone approach is strictly valid 
at $\gamma \to \infty$ and in this limit 
enables one to solve the problem 
of electron loss (ionization) 
and excitation exactly. One should also mention that  
the case of a strong ion-atom interaction   
this exact solution does not coincide 
with the first Born results, no matter how high 
is the impact energy \cite{abv-book}.   
Taking the above two points into account 
we can make the following conclusions. 
First, in the high-energy limit 
the SE model yields 
an exact solution for the transition amplitude 
of the projectile-electron excitation and loss. 
Second, even at $\gamma \to \infty$ the results 
of the SE model still in general differ 
from those of the first Born approximation. 

It is rather obvious that such  
conclusions would also hold 
if the electron transitions in the ion 
are caused by the collision with  
a charged particle (e.g. a stripped atomic nucleus) and, 
for instance, can be applied to $K$-shell ionization 
of atoms by high-energy bare nuclei.   
In this respect one should note that in the literature 
on relativistic ion-atom collisions 
there had been already attempts 
to consider the high-energy limit of 
distorted-wave models for the case 
of atomic ionization or excitation by 
the impact of a nucleus, where  
these processes are treated as 
a three-body Coulomb problem 
(the incident and atomic nuclei and atomic electron).  
In particular, starting with the work  
of \cite{deco-gruen-1}-\cite{deco-gruen-3}, 
it had been assumed that the high-energy limit 
of such distorted-wave models for ionization  
and excitation processes, 
like the continuum-distorted-wave-eikonal-initial-state 
(CDW-EIS) (see e.g. \cite{crothers}, 
\cite{vn-rdw-ioniz}) and the SE,  
is simply that of the first Born approximation. 

We have just seen, however, that 
for the SE model this is not true and that in 
the high-energy limit the transition amplitude, 
obtained in this model,  
goes over into the amplitude derived 
in the light-cone approach. 
Since at $\gamma \gg 1$ the CDW-EIS 
model becomes essentially 
identical to the SE,  
the high-energy limit of the CDW-EIS 
coincides with the light-cone approach 
but differs from 
the first Born approximation. 

\subsection{ The explicit form of the amplitude } 

Using the explicit form (\ref{e7}) 
of the scalar potential we obtain that   
\begin{eqnarray}
\int_{-\infty}^{+\infty} dt  
\exp(i \omega_{n0} t) \Phi(t) &=&  
\frac{ 2Z_A }{ v } %
\exp\left(i \omega_{n0} z/v \right) %
\sum_j A_j K_0\left( s_{\perp} \Lambda_j \right),  
\label{eic-new-6} 
\end{eqnarray} 
where where $K_0$ is the modified Bessel
function \cite{IStegun}, 
$s_{\perp} = | {\bf s}_{\perp}| = |{\bf r}_{\perp} - {\bf b}|$ 
and 
\begin{eqnarray}
\Lambda_j = \sqrt{\kappa_j^2 + \omega_{n0}^2/(\gamma^2 v^2)}.    
\label{eic-new-7}  
\end{eqnarray} 
Besides, it also follows from (\ref{eic-new-6}) that    
\begin{eqnarray} 
\int_{-\infty}^{+\infty} dt  \Phi(t) = 
\frac{ 2Z_A }{ v } %
\sum_j A_j K_0\left( \kappa_j \,  s_{\perp} \right).  
\label{eic-new-8}  
\end{eqnarray} 

Taking Eqs.(\ref{eic-new-6}) and (\ref{eic-new-8})  
into account the transition amplitude becomes   
\begin{eqnarray}
a_{fi}({\bf b}) &=& i \frac{ 2 Z_A c }{ v^2 }  %
\sum_j A_j  \left\langle \psi_n \left| \exp\left(i \frac{ \omega_{n0} z}{v} \right) %
\exp\left( i \frac{ 2Z_A }{ v } %
\sum_j A_j K_0\left(\kappa_j \, s_{\perp} \right) \right) \times 
\right. \right.   
\nonumber \\ 
&& \left. \left. 
\left( \frac{ \alpha_z }{\gamma^2} 
K_0\left( s_{\perp} \Lambda_j \right) 
- i \frac{ v \Lambda_j }{ \omega_{n0} s_{\perp}}  
K_1\left( s_{\perp} \Lambda_j \right) \, \,  
{\bf s}_{\perp} \cdot \mbox{\boldmath$\alpha$}_{\perp} \right) 
\right| \psi_0 \right\rangle,     
\label{eic-new-9}  
\end{eqnarray} 
where $K_1$ is the modified Bessel
function \cite{IStegun}.

\subsection{ The limit of vanishing screening }  

The results obtained above, can 
also be applied for treating 
ionization and excitation of neutral atoms 
in relativistic collisions with bare nuclei. 
This can be done by setting in (\ref{eic-new-9}) 
$\kappa_j=0$, replacing there 
$Z_A$ by $Z_N$, where $Z_N$ is the charge of 
the nucleus incident on the atom, 
and regarding now $\psi_0$ and $\psi_n$
as the initial and final states of the 'active' 
electron of the atom. Keeping in mind that 
$\sum_j A_j =1$, taking into account that 
$ K_0(x) \approx - \ln \left( \frac{x}{2} \right) - \Gamma $
for $\mid x \mid \ll 1$ (see e.g. \cite{IStegun}),
where $\Gamma $ is Euler's constant,  
and disregarding an inessential phase-factor 
we obtain     

\begin{eqnarray}
a_{fi}({\bf b}) &=& i \frac{ 2 Z_N c }{ v^2 }  %
\left\langle \psi_n \left| \exp\left(i \frac{ \omega_{n0} z}{v} \right) %
\exp\left( - i \frac{ 2Z_A }{ v } \ln s_{\perp} \right) \times 
\right. \right.   
\nonumber \\ 
&& \left. \left. 
\left( \frac{ \alpha_z }{\gamma^2} 
K_0\left( \frac{ \omega_{n0}}{\gamma v} s_{\perp} \right) 
- i \frac{ v \Lambda_j }{ \omega_{n0} s_{\perp}}  
K_1\left( s_{\perp} \Lambda_j \right) \, \,  
{\bf s}_{\perp} \cdot \mbox{\boldmath$\alpha$}_{\perp} \right)
\right| \psi_0 \right\rangle.    
\label{eic-new-10}   
\end{eqnarray} 

\section{ Applications }  

The SE model is normally regarded 
as a tool for describing collision-induced transitions 
between bound states. In our case it would mean that 
this model should be first of all applied   
for treating the projectile-electron excitation. 
It is known \cite{vnu-msea-excit} 
that at comparatively low relativistic 
impact energies the model has a problem with describing 
transitions involving electron spin flip. However, 
this problem diminishes when the energy increases 
and practically disappears at energies 
$\stackrel{>}{\sim} 1$ GeV/u. 

The SE model can also be used for considering 
the projectile-electron loss (in particular, the total cross section). 
In such a case the model also becomes more 
accurate when the impact energy increases.  
In particular, provided in the rest frame of the atom 
the magnitude of the velocity 
of the electron, emitted by the projectile, 
is of the order of the collision velocity, the SE model 
can be applied for calculating not only the total 
but also differential loss cross sections. 
In practical terms this condition holds starting 
already with $\gamma \simeq 2$--$3$, i.e. 
at impact energies $\sim 1$--$2$ GeV/u. 

As was shown in the previous section, 
at asymptotically high impact energies 
the SE model yields an exact solution. 
In the case of the projectile-electron excitation and loss 
in collisions with neutral atoms 
even for the most highly charged projectiles 
the region of such energies 
is actually reached already at $\gamma \sim 30$--$50$.  
Thus, the model should work excellently  
starting with 
the magnitude of $\gamma$ of a few tens. 

Taking all this into account 
one can expect that the SE model performs 
quite well at impact energies $\sim 1$--$30$ GeV/u,  
which are of special interest for the present study.  
Below the projectile-electron excitation and loss 
will be considered in this energy range 
for collisions between hydrogen- and helium-like 
highly charged ions and heavy atoms  
by using the SE model. Results 
of this model will also be compared 
with those of the first Born approximation. 
  
\subsection{ Single-electron loss } 
  
\begin{figure}[t] 
\vspace{-0.25cm}
\begin{center}
\includegraphics[width=0.45\textwidth]{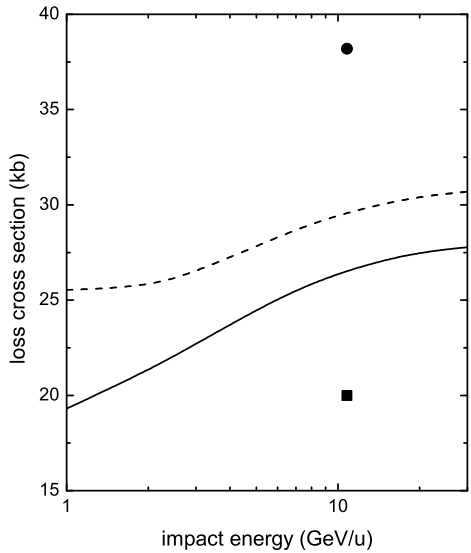} 
\end{center}
\vspace{-0.75cm}
\caption{ Cross section for the electron loss 
from Au$^{78+}$(1s) ions in collisions 
with neutral Au atoms.  
Solid curve: eikonal results. 
Dash curve: first Born results.   
The circle displays experimental data from 
\cite{high-0} while the square shows the result 
of \cite{high-1} scaled to the gold target. } 
\label{Au_s} 
\end{figure} 

In figure \ref{Au_s} we show results for 
the electron loss from incident Au$^{78+}$(1s)  
projectiles in collisions with neutral Au atoms at 
impact energies $1$-$30$ GeV/u. 
These results include our first Born and eikonal 
calculations as well as experimental 
data from \cite{high-0} and \cite{high-1} 
on the electron loss from 
$10.8$ GeV/u Au$^{78+}$(1s) projectiles.  

In the case considered our eikonal and first 
order results differ by $15$-$35\%$. As expected, 
when the impact energy increases the difference 
between them decreases. On overall, the difference 
is not large but nevertheless 
should be taken into account if precise cross section 
values are needed. 

Both the first Born and eikonal cross sections 
agree neither with the experimental data of \cite{high-0} 
nor with that of \cite{high-1}. The data from \cite{high-0} 
are substantially smaller (by about of $30\%$--$50\%$ ) 
while the data reported in \cite{high-1} 
are considerably larger (by about a factor of $1.3$--$1.4$) 
than our results \cite{explan1}. 

As was already mentioned in the Introduction, 
there is a difference by roughly a factor of $2$ 
between experimental cross sections reported for 
for the loss from $160$ GeV/u Pb$^{81+}$ ions in 
collisions with solid and gas targets with 
the solid state cross sections being larger.   
The origin of this difference, which was explained 
in \cite{we}, lies in multiple collisions between 
the projectile and atoms inside solids which effectively 
enhance the electron loss process. Since the magnitude 
of this difference depends on the impact energy and 
decreases when the energy decreases,  
similar reasons are probably responsible for 
the observed disagreement between 
our results for atomic targets and the 
experimental data from \cite{high-1}. 

\subsection{ Two-electron transitions  } 

If a projectile-ion initially carries several 
electrons, then more than one electron of the ion 
can be simultaneously excited and/or lost in
a single collision with a neutral atom. 
Helium-like ions are simplest projectiles, 
for which simultaneous transitions of more than 
one electron are possible, and below 
we consider double electron loss and loss-excitation 
for the case of such projectiles.  
  
It is known (see e.g. \cite{abv-book}) 
that, provided the condition $\frac{Z_I Z_A}{v} > 0.4$ 
is fulfilled, two-electron transitions 
in a heavy helium-like ion occurring in collisions 
with an atom are governed practically solely by 
the independent interactions between 
the atom and each of the electrons of the ion. 
In order to describe such transitions  
one can apply the independent electron model. 
According to this model the cross section 
for the double electron loss from 
a helium-like ion is given by 
\begin{eqnarray} 
\sigma_{l,l} = 2 \pi \int_0^{\infty} db b P_{l,l}(b),  
\label{ind-1} 
\end{eqnarray} 
where the probability $P_{l,l}(b)$ for 
the two-electron loss is given by  
\begin{eqnarray} 
P_{l,l}(b) = P_{loss}^2(b), 
\label{ind-2} 
\end{eqnarray} 
where $P_{loss}(b)$ 
is the single-electron 
loss probability  
in a collision with a given value of 
the impact parameter $b$. 

The cross section for the simultaneous 
loss-excitation in the case of a helium-like 
ion is evaluated as 
\begin{eqnarray} 
\sigma_{e,l} = 2 \pi \int_0^{\infty} db b P_{e,l}(b),  
\label{ind-3} 
\end{eqnarray} 
where the probability $P(b)$ for 
the two-electron process is given by  
\begin{eqnarray} 
P_{e,l}(b) = 2 \, P_{exc}(b) \, P_{loss}(b),    
\label{ind-4} 
\end{eqnarray} 
where $P_{exc}(b)$ is 
the single-electron excitation 
probability. 

\subsubsection{ Double-electron loss }

\begin{figure}[t] 
\vspace{-0.25cm}
\begin{center}
\includegraphics[width=0.45\textwidth]{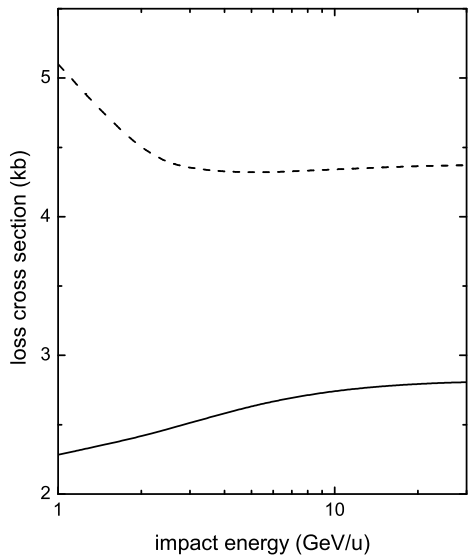} 
\end{center}
\vspace{-0.75cm} 
\caption{ Cross sections for double electron loss 
from Au$^{77+}$(1s$^2$) ions in collisions 
with neutral Au atoms.   
Solid curve: eikonal results. 
Dash curve:  first Born results. } 
\label{Au_d} 
\end{figure} 

Figure \ref{Au_d} shows 
calculated cross sections for double electron loss 
from Au$^{77+}$(1s$^2$) projectiles incident on neutral 
Au atoms at impact energies $1$--$30$ GeV/u. Compared 
to the single loss, now the difference between 
the eikonal and first Born results 
is much more pronounced. For impact energies 
$\sim 1$ -- $5$ GeV/u these calculations predict even 
qualitatively different dependencies of the cross sections 
on the collisions energy. 
The absolute difference between the first Born and eikonal 
cross sections ranges between $\simeq 2.5$ at $1$ GeV/u 
to $\simeq 1.5$ at $30$ GeV/u. Moreover, as  
additional calculations show, for impact energies 
above $30$ GeV/u the ratio of $\simeq 1.5$ remains 
almost a constant and, thus, even at asymptotically 
high collision energies the first Born calculation 
still substantially overestimates  
the cross section values. 

\subsubsection{ Simultaneous loss-excitation } 

\begin{figure}[t] 
\vspace{-0.25cm}
\begin{center}
\includegraphics[width=0.41\textwidth]{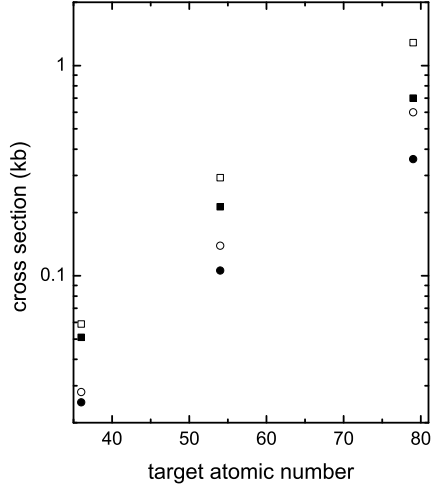} 
\end{center} 
\vspace{-0.75cm} 
\caption{ Cross sections for the simultaneous 
electron loss-excitation,  
U$^{90+}$(1s$^2$) $\to$ U$^{91+}$($n=2, j$) + e$^-$,      
occurring in collisions with Kr, Xe and Au 
atomic targets at $20$ GeV/u. 
$j=1/2$ and $j=3/2$ are the angular momentum 
of the states of the hydrogen-like uranium ion.  
Squares and circles show results for $j=1/2$ and $j=3/2$, respectively. 
Open and solid symbols denote the cross sections obtained using   
the first Born approximation and the SE model, respectively. }
\label{U_ion_exc} 
\end{figure}

\begin{figure}[t] 
\vspace{-0.25cm}
\begin{center}
\includegraphics[width=0.41\textwidth]{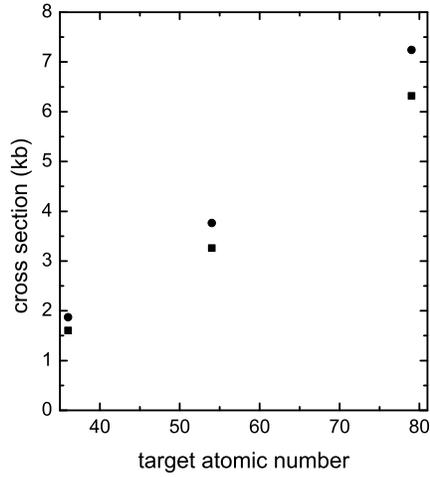}
\end{center} 
\vspace{-0.75cm} 
\caption{ Cross sections for excitation of  
$20$ GeV/u U$^{91+}$(1s) projectiles into 
the states with $n=2, j=1/2$ (squares) and $n=2, j=3/2$ (circles)  
in collisions with Kr, Xe and Au atomic targets.   
The cross sections were calculated using the SE model. } 
\label{U_excitation}
\end{figure}

In figure \ref{U_ion_exc} we present results 
of calculations for the simultaneous 
projectile-electron excitation and loss,     
U$^{90+}$(1s$^2$) $\to$ U$^{91+}$($n=2, j$) + e$^-$, 
where $n$ and $j$ are, respectively, the principal 
quantum number and the total angular momentum 
of the excited state of the hydrogen-like ion. 
The projectile is assumed to collide  
with Kr, Xe and Au atomic targets at 
an impact energy of $20$ GeV/u.   
Similarly to the case of the double electron loss, 
considered above, we observe that the differences 
between the cross sections, calculated with 
the first Born and eikonal probabilities, 
can be quite substantial if the atom 
is sufficiently heavy. 
For collisions leading to the population of 
the states with $j=1/2$ the first Born results are 
larger by a factor of 
$\simeq 1.16$ (Kr), $\simeq 1.38$ (Xe) and $\simeq 1.84$ (Au).       
When the states with $j=3/2$ are populated 
this ratio is $\simeq 1.12$, 
$\simeq 1.31$ and $\simeq 1.67$, respectively.  

We see that the difference between the first Born  
and eikonal cross sections turns out to be somewhat 
smaller for transitions involving the states with $j=3/2$.   
This can be attributed to the fact that, 
compared to the $j=1/2$ case, these transitions 
are characterized by larger impact parameters. 
As a result, the field of the atom 
acting on the electrons of the ion is weaker 
and, therefore, the first Born treatment 
becomes less inaccurate. 

As additional calculations show, 
the differences between the results 
of the eikonal and first Born calculation 
does not substantially change
when the impact energy increases further.  
Thus, like for the double electron loss, 
even at asymptotically high impact energies 
the first Born calculation may considerably 
overestimate the cross section for 
the simultaneous projectile-electron excitation and loss. 

\begin{figure}[t] 
\vspace{-0.25cm}
\begin{center} 
\includegraphics[width=0.41\textwidth]{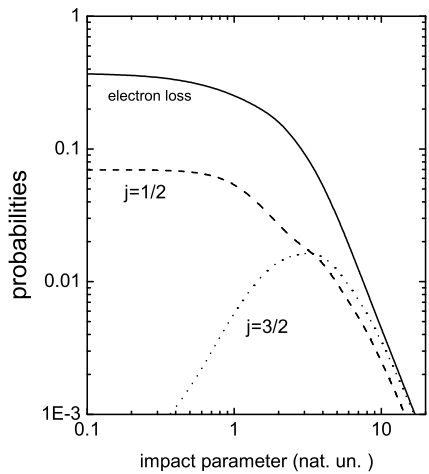}
\end{center} 
\vspace{-0.75cm} 
\caption{ Probabilities for the electron loss, 
excitation to the states with $(n=2$, $j=1/2)$ and  
$(n=2$, $j=3/2)$ in collisions of 
$20$ GeV/u U$^{91+}$(1s) ions with Au atoms. }
\label{U_prob}
\end{figure}

According to both the first Born and eikonal results 
the cross sections for the simultaneous loss-excitation 
are larger for transitions to the states with $j=1/2$. 
However, the pure excitation U$^{90+}$(1s) $\to$ U$^{90+}$(1s; n=2, j) 
(or the excitation U$^{91+}$(1s) $\to$ U$^{91+}$(n=2, j), 
see figure \ref{U_excitation}) at these rather high energies 
proceeds already more efficiently into the states with $j=3/2$. 
The origin of this interesting peculiarity can be understood 
by considering the single-electron transition probabilities. 
It turns out that the probability for the electron loss has a larger 
overlap with the probability for the excitation 
into the states with $j=1/2$ (see for illustration figure 
\ref{U_prob}). 

\section{Conclusions} 

We have considered the symmetric eikonal model 
for treating projectile-electron transitions 
in relativistic collisions with heavy atoms. 

We have shown that at asymptotically 
high impact energies this model yields  
the same results as the light-cone approach 
and, thus, offers an exact solution 
for the transition amplitude.  
In the limit $\gamma \gg 1$  another 
popular distorted-wave model -- 
the continuum-distorted-wave-eikonal-initial-state 
approximation -- becomes essentially 
identical to the symmetric eikonal model. 
Therefore, at asymptotically high impact energies the CDW-EIS 
also becomes exact and, contrary to what was stated earlier, 
does not coincide with the first Born approximation. 

Using the SE model and 
the first Born approximation 
we have calculated cross sections 
for the projectile-electron excitation and loss. 
We have shown that at impact energies $ \sim 1 $--$ 30 $ GeV/u, 
which are relevant for the future GSI facility, 
there are noticeable deviations between the eikonal and 
first Born results for single-electron transitions 
and very substantial differences between such results for 
transitions involving two projectile electrons. 
Moreover, these very substantial differences 'survive' 
even in the asymptotic limit $\gamma \to \infty$.   

Our results clearly show a great advantage 
of the SE model over the first Born approximation. 

\section*{Acknowledgment}

B.N. acknowledges the support from the 
Deutsche  Forschungsgemeinschaft under the project VO 1278/2-1.


\begin{thebibliography}{99}

\bibitem{standf} R. Anholt, W.E. Meyerhof, X.-Y. Xu, 
H. Gould, B. Feinberg, R.J. McDonald, H.E. Wegner and 
P. Thieberger, Phys. Rev. {\bf A 36} 1586 (1987); 
W.E. Meyerhof, R. Anholt, X.-Y. Xu, 
H. Gould, B. Feinberg, R.J. McDonald, H.E. Wegner and 
P. Thieberger, NIM {\bf A 262} 10 (1987)  

\bibitem{low-middle-1} Th. St{\"o}hlker et al, 
Nucl. Instr.Meth. {\bf B 124} 160 (1997) 

\bibitem{low-middle-2} C. Scheidenberger and H. Geissel, 
Nucl. Instr.Meth. {\bf B 135} 25 (1998) 

\bibitem{low-middle-3} C. Scheidenberger, Th. St{\"o}lker, 
W.E.Meyerhof, H. Geissel, P.H. Mokler and B. Blank, 
Nucl. Instr.Meth. {\bf B 142} 441 (1998) 

\bibitem{St3} T. Ludziejewsky, T.St\"ohlker, 
C.D.Ionesku, P.Rymuza, H.Beyer, F.Bosch, 
C.Kozhuharov, A.Kr\"amer, D.Liesen, 
P.H.Mokler, Z.Stachura, P.Swiat, 
A.Warczak, R.W.Dunford, 
Phys. Rev. {\bf A 61} 052706 (2000)  

\bibitem{high-2} H.F. Krause, C.R. Vane,
S. Datz, P. Grafstr{\"o}m,
H. Knudsen, S. Scheidenberger,
and R.H. Schuch, Phys. Rev. Lett.
{\bf 80}, 1190 (1998)

\bibitem{high-3} H.F. Krause, C.R. Vane,
S. Datz, P. Grafstr{\"o}m, H. Knudsen,
U. Mikkelsen, S. Scheidenberger, 
R.H. Schuch, and Z. Vilakazi, 
Phys. Rev. {\bf A 63} 032711 (2001) 
 
\bibitem{high-4} C.R.Vane and H.F.Krause,  
NIM {\bf B 261} 244 (2007) 

\bibitem{high-0} A. Westphal and Y.D.He, 
Phys.Rev.Lett. {\bf A 71} 1160 (1993) 

\bibitem{high-1} N. Claytor, A. Belkacem, 
T. Dinneen, B. Feinberg, 
and H. Gould, Phys.Rev. {\bf A 55} R842 (1997) 

\bibitem{we} A.B. Voitkiv, B. Najjari, and A. Surzhykov, 
J.Phys. {\bf B 41} 111001 (2008) 

\bibitem{mont-review} E.C. Montenegro, W.E. Meyerhof
and J.H. McGuire, Adv.At.Mol. and Opt. Phys. 
{\bf 34} 249 (1994)

\bibitem{abv-book} A. B. Voitkiv and J. Ullrich, 
{\it Relativistic Collisions of Structured Atomic Particles} 
(Springer-Verlag, Berlin, 2008) 

\bibitem{ascr} According to first Born considerations,  
in the antiscreening mode the nucleus of the atom 
does not at all interact with the electron of the projectile. 

\bibitem{Moliere} G. Moliere, 
Naturforsch. {\bf 2A} 133 (1947) 

\bibitem{Salvat} F. Salvat, J.D. Martinez, 
R. Mayol, and J. Parellada, 
Phys.Rev.{\bf A 36} 467 (1987)  

\bibitem{baltz} A.J.Baltz, Phys.Rev.Lett. {\bf 78}, 
1231 (1997). 

\bibitem{deco-gruen-1} G.R.Deco and N.Gr\"un, 
J.Phys. {\bf B 22} 1357 (1989) 

\bibitem{deco-gruen-2} G.R.Deco, K.Momberger 
and N.Gr\"un, J.Phys. {\bf B 23} 1990 (1989) 

\bibitem{deco-gruen-3} G.R.Deco and N.Gr\"un, 
J.Phys. {\bf B 22} 3709 (1989) 

\bibitem{crothers} D.S.F. Crothers,  
{\it Relativistic Heavy-Particle 
Collision Theory}, 
Kluwer Academic/Plenum Publishers, 
London (2000)  

\bibitem{vn-rdw-ioniz} A.B.Voitkiv and B.Najjari, 
J.Phys. {\bf B 40} 3295 (2007)  

\bibitem{IStegun} M. Abramowitz and I. Stegun,
{\it Handbook of Mathematical Functions}
(Dover Publications, Inc., New York, 1965)




\bibitem{vnu-msea-excit} A.B.Voitkiv, B.Najjari 
and J.Ullrich, Phys.Rev. {\bf A 75} 062716 (2007)  

\bibitem{explan1}   
Note that the experimental data reported in  
\cite{high-1} are in a very good agreement 
with results of theoretical considerations 
of \cite{anh-beck} and \cite{as}. 
However, the theoretical models of \cite{anh-beck} 
and \cite{as}, which were developed 
for collisions between individual ions 
and atoms (and, therefore, should be first 
of all tested in collisions with gas targets), 
overestimate experimental data obtained in collisions 
with rarefied gas targets \cite{high-3} 
by a factor of $2$-$3$ (while our results agree 
very nicely with this atomic target experiment). 
The critical discussion of the models 
of \cite{anh-beck} and \cite{as} can be found 
in \cite{abv-book}.     

\bibitem{anh-beck} R. Anholt and U. Becker,  
Phys.Rev. {\bf A 36} 4628 (1987)

\bibitem{as} A.H. S{\o}rensen, Phys.Rev. 
{\bf A 58} 2895 (1998) 


\end{thebibliography}
\end{document}